# Mars Missions Failure Report Assortment Review and Conspectus

Malaya Kumar Biswal M[†] and Ramesh Naidu Annavarapu[‡]

*Department of Physics, School of Physical Chemical and Applied Sciences,*

*Pondicherry University, Kalapet, Puducherry – 605 014, India*

**Mars has been successfully explored by various space firms. Globally, 44 mission attempts were made, lamentably 26 encountered setbacks. There have been instances where a trivial error in the progressive accomplishment of spaceflight sequence might prompt in an extreme loss. Hence in this paper, we have investigated several failure records and investigation reports of spacecraft attempted towards Mars. The report comprises a precise and summarized report assessed from various adequate online resources as well as published research articles. Analyzing records, spaceflight sequence charts, and mission targets achieved were technically tabularized beginning from launch to the Mars landing. Additionally, we have estimated approximate mission duration, the rate of failure, and their proportions and were graphically portrayed. Further, we have generated graphical representations of the number of spacecraft launches concerning the country, type of spacecraft, and launch vehicles.**

## I. Nomenclature

| | | |
|------|---|-----------------------------|
| *AD* | = | Aerodynamic Deceleration |
| *BSS* | = | Booster Stage Separation |
| *CRC* | = | Communication Relay Check |
| *CSS* | = | Cruise Stage Separation |
| *DS* | = | Deep Space |
| *EDL* | = | Entry, Descent, and Landing |
| *EDM* | = | Entry, Descent Module |
| *ESA* | = | European Space Agency |
| *HBE* | = | Hypersonic Ballistic Entry |
| *HSJ* | = | Heat Shield Jettison |
| *IMU* | = | Inertial Measurement Unit |
| *LDNG* | = | Landing (Ground Touchdown) |
| *LEO* | = | Low Earth Orbit |
| *LEOI* | = | Low Earth Orbit Insertion |
| *LNH* | = | Launch |
| *MA* | = | Mars Approach (Arrival) |
| *MCO* | = | Mars Climate Orbiter |
| *MMH* | = | Mono Methyl Hydrazine |
| *MPL* | = | Mars Polar Lander |

[†] Graduate Researcher, Department of Physics, Pondicherry University, India; malaykumar1997@gmail.com, mkumar97.res@pondiuni.edu.in, Member of Indian Science Congress Association, Student Member AIAA
[‡] Associate Professor, Cognitive Neuroscience Expert, Department of Physics, Pondicherry University, India; rameshnaidu.phy@pondiuni.edu.in, arameshnaidu@gmail.com, Non-Member AIAA

| MV | = | Mars Venus |
|-----|---|---|
| NASA | = | National Aeronautics Space Administration |
| NTO | = | Nitrogen Tetroxide |
| MOI | = | Mars Orbital Insertion |
| OIB | = | Orbital Insertion Burn |
| OMB | = | Orbital Maneuver Burn |
| PDD | = | Parachute Deployment Descent |
| RPD | = | Retrorocket Powered Descent |
| SB | = | Stage Burns |
| TCM | = | Trajectory Correction Maneuver |
| TMTI | = | Trans-Mars Trajectory Insertion |
| USA | = | United States of America |
| USSR | = | Union of Soviet Socialists Republic |
| UTC | = | Universal Time Coordinated |
| VSS | = | Vehicle Stage Separation |

## II.   Introduction

Exploration is one of the attentive endeavors to mankind and a strategy for evolution. We have been incessantly reconnoitering our neighboring planet and the universe since the twentieth century. The progression of rocketry and planetary science in the last decade engendered a futuristic window to explore the red planet which has been a source of inspiration to hundreds of space explorers. Globally, 44 mission attempts were made, lamentably 26 encountered setbacks. There have been instances where a trivial error in the progressive accomplishment of spaceflight sequence might prompt in the extreme loss and unsuccessful mission. However, striving a robotic spacecraft towards the red planet requisites a great extent of attentiveness, since no effective recovery procedures can be executed to recover the probe once it cruised to Mars than the probes stranded in LEO. So it significant to have a concise understanding of the source behind loss of Mars probes. Considering this complication, space agencies mobilize the mishap investigation review board to interpret and release the root basis responsible for the failure of the spacecraft. No reports were assorted unveiling the root causes behind earlier lost space probes. It is because many space firms never revealed the failure reports as part of their secret agenda. Hence in this paper, we scrutinized and recapitulate failure reports of all collapsed spacecraft directed towards Mars since 1960.

## III.   Research Methodology

### A. Research Methodology

- For our summarily report, failure records of the spacecraft launched between 1960 to 1973 were gathered for study and analysis from National Space Science Data Coordinated Archive, online web archives, NASA's Solar system server, NASA Technical Report Server, and adequate online sources.
- Records gathered from online resources were compared with other resources and published articles to verify the authentic analogy.
- For the spacecraft launched between 1988 to 2016, the reports were collected and investigated from research articles and appropriate web pages of space agencies.
- The reports were carefully analyzed and the inappropriate report described in various articles was precisely summarized.
- Subject to our study, spaceflight sequence chart was prepared to define the accomplishment of mission target along the mission sequence and graphical representations were made for the number of spacecraft launches to their respective launch vehicles, country, proportions of issues encountered with probes and the type of spacecraft.

# IV. Mars Flybys

## A. Mars 1M.No.1

The first-ever spacecraft attempted in the direction towards Mars to photograph the planet on a flyby trajectory that endured in failure. The failure was related to the third stage resonant vibration provoked by a faulty gyroscope that ultimately impaired the attitude control system of the launch vehicle (Molniya). Multitudinous vibrations spawned as a result of the synergy of other boosters with upper stage booster persecuted the flight. Sequentially, the horizon sensor detached from the booster and the launch vehicle nosedived from the usual flight path angle. As a consequence, the ground commanded third stage engine to halt engine burns posterior to five minutes into the flight, during this phase the spacecraft uplifted to an extent of 120km. Thereafter, it re-entered into the earth's atmosphere and destroyed in lower earth orbit [1-3].

## B. Mars 1M.No.2

The Soviet Union launched its second spacecraft predecessor to 1M No.1 as part of the "Mars Program" to explore Mars. Nevertheless, after (T+290) seconds into the flight, due to the leakage in the oxidizer shut- off valve, it made liquid oxygen spill around the engine's fuel inlet valve. This leakage ultimately froze the third stage engine fuel (kerosene) resulting in failed ignition of 80715K engine caused by shut-off of the third stage engine valve, following this issue the spacecraft reached an altitude of 120km above the earth surface. As a consequence, the spacecraft failed to achieve LEO and burned up in the earth's atmosphere [4-6].

## C. Mars 2MV-4 No.1

After two successive failures, Russians sent a new spacecraft (Sputnik-22) that successfully lifted-off from the launch pad aboard Molniya launcher to conduct Mars flyby. Directly after its launch, the block 'L' upper stage started to ignite. Then the lubricant leaked out of the turbopump and consequently made the main engine to explode and destroy the spacecraft [7]. According to a report [8], twenty-two pieces of spacecraft debris disintegrated and decayed between 29 October 1962 and 26 February 1963.

## D. Mars 2MV-4 No.2 (Mars 1)

Excluding four repeated failures, Russians re-attempted Mars 1 whose launch remained a success. After the fourth stage separation, the spacecraft left LEO and the solar panels were successfully deployed. Telemetry data indicated that the spacecraft transferred to a gyroscopic stabilization state due to the leakage of one of the gas from gas valves in the orientation control system. During this notch, sixty-one radio transmissions were achieved at the five-day interval. Following this issue, the ground lost the communication due to the failure of the spacecraft's orientation system on 21 March 1963 [9].

## E. Mariner 3

The first American spacecraft attempted in the vicinity of Mars. The Mariner 3 inquisition board reported that one hour after the launch, there was no indication of solar panel deployment and all the instruments were reported to be working properly. Telemetry data suggested that there was a problem in separation due to either launch vehicle or payload fairing. Later, it was identified that a protective heat shield failed to eject after the spacecraft had passed through the atmosphere. Following this, the ground commanded the spacecraft to jettison its heat shield but nothing happened. As a result, the spacecraft lost power and battery died due to un-deployment of solar panels. Power lack aboard probe affected the communication system leading to the termination of communication from the probe. Latterly due to its lower velocity, it failed to achieve the Trans-Mars trajectory path [7, 10].

## F. Zond 2

The prime reason behind the failure of the spacecraft was a failure in the deployment of solar panels during the voyage along the trans-Mars trajectory, caused due to the damage of a tug cord during the Block L upper stage separation from the rocket which was designed to pull and deploy the solar panels. Pursuing these concerns, the controllers were able to deploy the solar panels on 15 December 1964, but it was too late to perform midcourse maneuver correction to flyby Mars. Additionally, radiators of the thermal control system and programmed timer also affected during trans- interplanetary injection which led to an unsuited thermal condition of the spacecraft. This resulted in the loss of communication from the spacecraft [7, 11].

# V. Mars Landers

## A. Mars 2MV-3. No.1

Despite three failures, the Soviet Union repeatedly launched 2MV-3 No.1 onboard Molniya launch vehicle. Preliminary to 4 minutes and 33 seconds into the flight (T+260 sec). The oxidizer pressurization system malfunctioned causing cavitation within the turbo-pump feed lines at T+32 seconds. Despite this issue, the lower stage of the rocket delivered the payload to LEO. But the vibrations due to either cavitation or stage separation problem displaced the electrical controlling system of the ignition engine. Consequently, this obstructed the Block 'L' upper stage from igniting and leaving the spacecraft in parking orbit. Following these concerns, the spacecraft started to decay from the next day of its launch. The spacecraft debris remained in orbit until 19 January 1963 [12].

## B. Mars 2

Russia's first Mars probe to carry both orbiter and lander. The probe successfully approached Mars. But 4.5 hours before reaching Mars, the Mars 2 descent module separated from the orbiter on 27 November 1971. The descent module entered the Mars atmosphere relatively at 6 km/s. Following this phase, the lander unexpectedly malfunctioned and entered at a steep angle. EDL sequence did not occur as programmed and the parachute did not deploy. As a result, the lander made a great impact and crashed on the surface approximately at location 45◦S47◦E [13-15].

## C. Mars 3

The first artificial object to perform effective landing on any other planetary surface. After successful touchdown, the communication between earth stations and the lander module was established via Mars 3 orbiter. Approximately at 13:52:25 UTC (nearly 20 seconds after landing), the transmission ceased for unknown reasons, and no further communication was re-established. It is still uncertain whether the problem persisted in the lander or the communication relay on the orbiter. The lander malfunction is related to extreme Martian dust storms. These storms might have damaged the communication system thereby inducing coronal discharge [16-17].

## D. Mars 6

The Mars 6 became the second human-made object to effectuate successful landing on Mars. During the Mars transit eminently after the first mid-course correction on 13 August 1973, there was trouble in the telemetry system indicating difficulty in establishing communication. The problem was most likely to be caused by the effect of bad 2T312 transistor which was responsible for failure onboard computer of past Mars 4 orbiter. Despite the telemetry issue, the spacecraft operated autonomously and pursued its function as programmed. The lander separated from the flyby bus orbiter on 12 March 1974 and entered the Martian atmosphere. Subsequently, the parachute system deployed to cut down the terminal velocity. Preliminary to its precision landing, the ground controllers lost communication from the lander. Later, investigations estimated that due to its landing in geographically rough terrain, the radio communication system might have been damaged. Whatever, might be the reason for failure, the landers transmitted atmospheric data via Mars 6 telecommunication relay while descent [7, 15, 16, 18, 19].

## E. Mars 7

Mars 7 the fourth spacecraft of M-73 series successfully launched and inserted into Mars trajectory path. En route to Mars, it encountered communication issues and ground controllers were coerced to communicate via the radio communication system. On 9 March 1974, the landing module denied separation command from flyby bus but latterly separated. Consequently, the main retrorocket engine failed to ignite to initiate hypersonic atmospheric entry, but the failed ignition was identified due to the installation of a faulty transistor in onboard computer circuits. Finally, the entry vehicle missed the planet by 1,300 km and entered a heliocentric orbit [7, 15, 16, 19].

## F. Mars 96

Mars 96 was the heaviest spacecraft mission ever attempted in the 20th century as well as the only planetary probe of Soviet Russia in twelve years since Phobos mission. Rear to its launch on 16 November 1996, the carrier rocket Proton successfully placed the spacecraft into a parking orbit. But the Block D-2 fourth stage malfunctioned and failed to ignite. Consequently, the spacecraft re-entered the earth's atmosphere and crashed somewhere near Chile. Later on, the investigation team has failed to portray the exact reason behind Mars 96 fourth stage ignition failure due to a lack of telemetry data during missions [20-22].

## G. Mars Polar Lander

Mars Polar Lander or Mars Surveyor 98 lander was successfully launched and approached Mars. On 03 December 1999 after the cruise stage separation from the flyby bus, the lander module performed hypersonic atmospheric entry. At entry altitude, the antenna adverted off-Earth leading to the loss of communication from ground controllers. The prime cause of the loss of communication is ascertained. However, no signals were received from Mars Polar Lander as well as the Deep Space 2 probe [23-26].

The presumable factor for the loss of MPL is the unanticipated shutdown of the lander's retrorocket engine due to weird signals spawned through flawed MPL flight software in the interim of descent phase. The unauthentic signal would have indicated that the lander had landed before landing due to incorrect identification of vibrations provoked during the leg deployment phase. Consequently, the software persuaded the engine to shut down. The status of the lander is still uncertain due to the lack of flight data. It is difficult to predict whether the lander had touched down or crashed into the surface [27-29].

## H. Beagle 2

European Space Agency's made an excellent landing on Mars in their first attempt. After performing effective landing on 25 December 2003, Beagle 2 has contacted the 2001 Mars Odyssey but the ground controllers failed to receive signal. Several attempts were made to establish communication that remained ineffective. Eventually, no communication was ever re-established and declared lost on 6 February 2004 [30-33].

The fundamental cause for the loss of Beagle 2is still uncertain due to a lack of successful flight data from the lander module during EDL performance. Besides, it is very difficult to prognosticate the cause for failure. Hence, the Beagle 2 investigation board released two reports after six months of the internal investigation that summarizes two possibilities for the failure of the lander (i.e., technical and programmatic issue). In addition to this, several considerable factors such as robustness nature of air-bad design, inadequate testing Program, the possibility of collision between the back cover and the main parachute of lander module, premature deployment of the lander from the air-bag landing system, are also censurable for the loss of Beagle 2 lander [34-35].

## I. Fobos-Grunt

Fobos Grunt was Soviet Union's sample return mission cruised to moon Phobos. The probe Fobos-Grunt along with Yinghuo-1 (Chinese Mars Orbiter) uplifted onboard Zenit-2SB41 launch vehicle on 08 November 2011. Sequentially, Zenit injected the spacecraft into LEO, after successful orbiter insertion the scheduled cruise stage firing did not take place to propel the spacecraft towards Mars trajectory. The failed ignition was due to the malfunction of onboard computers considering a concurrent reboot of its two channels. The impairment of computers was either due to radiation damage of electronic chips or the installation of ill-equipped electronic components. The collapse of the onboard computer program due to the ruined chip made the spacecraft computer reboot persistently leaving the spacecraft stranded in low earth orbit. Eventually, the stage burn never occurred and the spacecraft was destroyed during re-entry [36-40].

## J. Schiaparelli EDM

European Space Agency's second attempt to land on Mars with the Schiaparelli demonstration module remained unsuccessful. The lander review board revealed that during landing attempt, ground controllers unexpectedly lost communication from the lander just one minute ahead of scheduled touchdown. Following communication failure, the lander performed automated landing. During entry, descent, and landing phase, the unexpected fluctuation in dynamics of the landing vehicle made the gyroscope (Inertial Measure Unit) incapable of calibrating higher readings. The failure of the gyroscope provoked fatal errors in the guidance and control system. Thus the EDL flight software, generated negative altitude data (below ground level) resulting in premature lander separation and hard impact onto the surface. Furthermore, considering factors such as inadequate enduring time of IMU, inadequate handling of IMU, inadequate design robustness, and contingency in hardware management are also accountable for the mishap of the lander [41-43].

## VI. Mars Orbiters

### A. Mars 2M.No.521

Soviet Union's M-69 series - a new generation spacecraft is primarily intended for studying Mars from orbit. Consequently, after launch especially after the first and second stage booster burns, the third stage ignition did not ordain on time. Several investigations reveal that the imbalance of a rotor in the third stage booster's oxidizer pump resulted in the loss of thrust and vehicle separation. Following this issue, the booster exploded and impacted in the mountains of Altai [44-46].

### B. Mars 2M.No.522

Similar to its predecessor attempts, Russians re-attempted M-69 spacecraft (2M No.522). Disparate studies conceded that, immediately after launch, the first stage engine of proton K/D UR-500 caught fire while lift-off. The fire was most likely to be caused by leakage of nitrogen tetroxide fostered by a lack of drain plug. Besides, this issue, remaining engines insisted stage burns to remunerate the flight for 30 seconds. But the thrust section went out of control and the rocket began to tilt horizontally before the engines were manually commanded to shut-off stage burns from their appropriate ground controllers. Eventually, the rocket nosedived into the ground covering the launch complex after 41 seconds into the flight [44-46].

### C. Mariner 8

The earlier investigation reported that the main cause for the failure of Mariner 8 was the failure of the entire guidance system during activation of the autopilot function. Subsequent analysis unveiled that, a diode equipped for protecting the spacecraft system from transient voltages was damaged during the replacement/ installation of a pitch amplifier circuit board which led to the launch vehicle malfunction and failed launch [7, 13, 47].

### D. Kosmos 419

This is the first of the fifth-generation spacecraft of the Soviet Union launched to overtake US Mars probes. After its launch, the vehicle successfully injected the spacecraft into a low earth parking orbit, then the Block D upper stage of the Molniya rocket failed to ignite as the ignition timer was incorrectly set. Later, the investigation showed that there was human error in programming eight-digit code to ignition timer. The timer had been set to ignite after 1.5 years instead of 1.5 hours to perform trans-mars trajectory maneuver. The result of which, the spacecraft re-entered and decayed in the upper atmosphere on 12 May 1971 just after two days of its launch [7, 48].

## E. Mars 4

Mars 4 was one of the 3M (M-73) spacecraft plighted to study Mars from the trajectory path. Succeeding its launch, the Proton's Block- D upper stage engine successfully placed the spacecraft into Trans-Mars trajectory path. After trajectory correction performed on 30 July 1973, two of three channels of onboard computers failed due to defective transistor which led to the malfunction of breaking engine plighted for second mid-course correction. As a result, the probe failed to achieve Mars orbit on 10 February 1974. Rather than its failures, ground controllers were able to command the spacecraft to transmit data, it transmitted radio occultation data and two panoramic surface images of Mars during the flyby [7, 15, 18, 19].

## F. Phobos 1

Soviet Union's 1988-Phobos 1 & Phobos 2 were acquisitive missions propelled towards the Martian moon (Phobos). On 1 September 1988 in transit to Phobos, Phobos 1 did not respond to multiple command requests indicating intricate in establishing communication with the ground controllers during the planned session. The investigation reported that at the interim of the communication session on 28 August 1988, a ground controller insensibly transmitted a wrong command to Phobos circumventing the proofread of computer which eventually turned-off the thrusters of attitude control system/stabilization system and orientation system. Ensuing this issue, Phobos 1 transposed its solar panel orientation away from the sun which readily discharged the battery leading to the loss of power required to respond to the powerful radio signals from the earth. Consequently, the communication from Phobos 1 was ceased terminating the mission strategies [7, 49-52].

## G. Phobos 2

Phobos 2 was a partially successful mission, on 27 March 1988 after changing its orientation to image Phobos, it encountered radio communication loss. Several attempts were made to re-establish radio contact that remained unsuccessful. Four hours later, ground controllers received a weak signal indicating the spacecraft spinning in off-design mode and lost all its orientation that adversely affected the spacecraft system from generating power. The main cause of the failure of Phobos 2 was again due to the failure of the orientation system due to simultaneous malfunction in both channels of onboard spacecraft computers [7, 49, 53].

## H. Mars Observer

Seventeen years after the Viking Program, the US launched Mars Observer for detailed scientific observation of Mars. The probe completed an interplanetary cruise to Mars. On 21 August 1993, three days before Mars orbital insertion, Mars Observer lost communication from ground controllers significantly due to problem emerged as a result of inappropriate pressurization of rocket thruster fuel tanks. Several attempts were made to re-establish the communication but the attempts remained unsuccessful. Lattery, the extensive analysis revealed that the major reason for the loss of spacecraft was due to the rupture of the fuel tank provoked under improper fuel pressurization of the propulsion system onboard spacecraft resulting in the exhalation of liquid monomethyl hydrazine and helium gas beneath spacecraft's thermal blanket. The leakage was endorsed as a result of inadvertent mixing of nitrogen tetroxide (NTO) and monomethyl hydrazine (MMH) in pressurized titanium tube during helium pressurization and their reactions have ruptured the tubing system. The unsymmetrical leakage of fuel made the spacecraft spin at a higher rate which adversely affected the transmitter switching and solar arrays orientation resulting in the expeditious discharge of batteries and loss of power. In addition to this issue, the leaked monomethyl hydrazine impaired the electrical circuits onboard spacecraft [54,55].

Moreover, multiple factors such as impairment of electrical power system caused as a result of short circuit of regulated power bus, failure of fuel tank pressurization regulator, rapid expulsion of NASA Standard initiator from a pyro valve that damaged the fuel tank, failure of computation function of spacecraft and failure of transmitters were also related to the loss of spacecraft [56-58].

## I. Nozomi

Japan's first step to explore Mars began with the launch of Nozomi (Planet-B) on 03 July 1998. After its successful launch, Nozomi performed powerful gravitational pull on 20 December 1998 due to defective thrust valve following two lunar gravity assist on 24 August and 8 December 1998 thereby traveling 1000 km. During this critical stage, the spacecraft consumed excess fuel than anticipated. Following this issue, Nozomi effectuated two earth gravity assists to propel itself in a trajectory towards Mars. Ultimately the electrical system and the S-band communication system were imparted by the solar eruption in April 2002 that provoked communication issues with the spacecraft. Moreover, the failure of the electrical system affected the thermal control system which solidified spacecraft propellant required for maneuvering. Subsequent attempts were made to heat the frozen propellant with solar radiation that remained ineffective. On 9 December 2003, the Nozomi team failed to rectify the trajectory maneuver after repeated attempts and concluded to terminate the mission. Afterward, the controllers cruised off the spacecraft to heliocentric orbit to avoid impact with other Marscrafts [59-62].

## J. Mars Climate Orbiter

The United States' last orbiter mission of the 20th century, Mars Climate Orbiter was successfully launched and intended to study the Martian climate. However, the probe failed before Mars orbital insertion. The Mars Climate Orbiter's Mishap Investigation Board obligated that the core reason for the loss of spacecraft was the failure in utilizing metric units [63-68]. The thruster performance data was to be in SI (metric) units rather than English units in a software file entitled "Small Forces". As a result, the Mars Surveyor Operation Project's System Interface Specification software was instructed to use thrust units as pounds-seconds (lbf-s) instead of Newton- seconds (N-s) which led to the computation of spurious trajectory path. Consequently, the spacecraft entered the Martian atmosphere at a lower altitude resulting in the destruction of spacecraft in the upper atmosphere or re-entered into a heliocentric orbit. Additionally, untraveled changes in spacecraft velocity, anomalous nature of navigation team with the spacecraft, interruption of $5^{th}$ trajectory maneuver correction, inadequate system engineering process, the improper link between project elements, lack of navigation team staffing, and training including faulty verification and validation process were also considerable factors for loss of Mars Climate Orbiter spacecraft [69-73].

## K. Yinghuo-I

Yinghuo-1 was the first Chinese interplanetary spacecraft intended to detect and observe the Martian magnetosphere and ionosphere [74]. This spacecraft was found to be lost along with Russian's Fobos-Grunt mission on 15 January 2012 and its disintegrated parts fell over the Pacific Ocean [75-76].

## VII. Mars Rovers

## A. Prop-M Rover

Both Mars 2 and Mars 3 lander had 4.5 kg Prop-M rover along with two penetrometers intended to measure the density of Martian soil. However, one rover lost with Mars 2 lander crash and another rover with Mars 3 lander which was never deployed on the surface [77].

## VIII. Results and Discussions

Analyzing overall spacecraft records, we have precisely summarized the root causes behind every spacecraft's failure. The report distinctly portrays that the spacecraft launched between 1960 and 1996 has concerns with the function of launch vehicles that have been sorted out nowadays. It contributes 26% (where launcher encounters issues with stage ignition and payload fairing) and 12% (where launcher experiences complete malfunction) shown in Fig-1 [81]. But the spacecraft attempted after 1996 had technical concerns (software and programmatic) that have to be taken into cogitation for prospects (i.e., it includes adequate testing of software program, fabrication of robust computers, and advanced communication system). The second failure proportion attributes to Software and Programming and third most 9% attribute to impairment of communication system shown in Fig-1. From the theoretical records, we have generated graphical presentation of proportions of spacecraft issues (Fig-1), approximate mission duration from the date of launch to the end of the mission (Fig-2) and number of spacecraft launches by type

Proportions of Issues Estimated

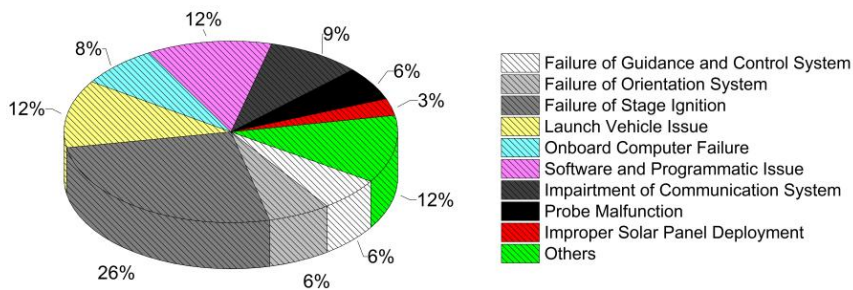

**Fig. 1 Estimated proportions of Issues [79]**

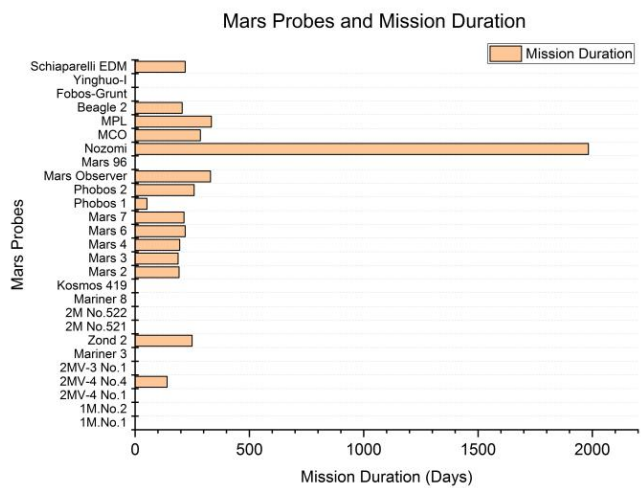

**Fig. 2 Approximate mission duration of Probes [80]**

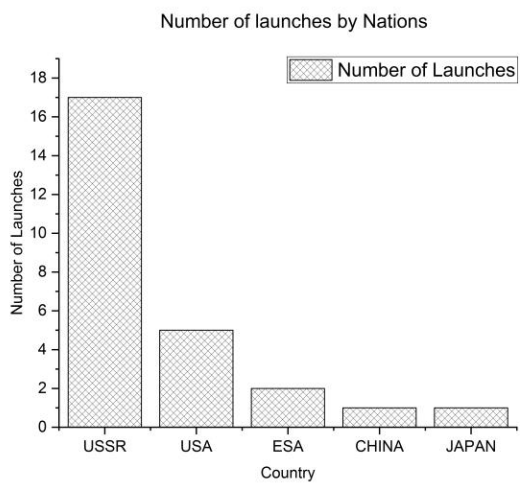

**Fig. 3 Number of Probes failed by Country**

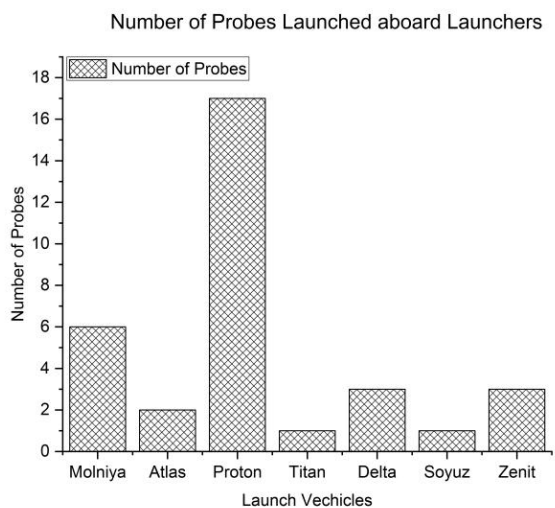

**Fig. 4 Number of probes launched aboard launchers.**

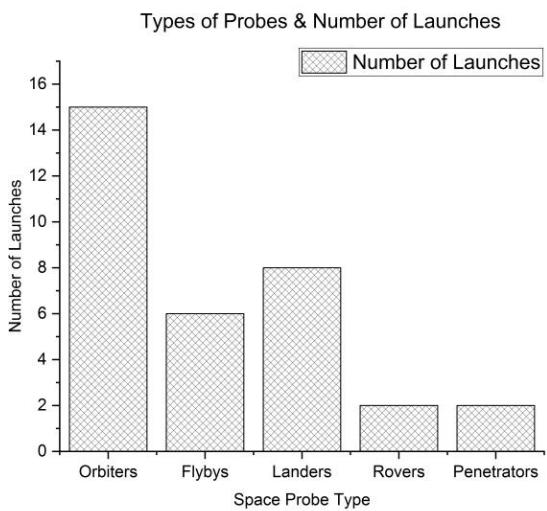

**Fig. 5 Number of Failed probes by types**

## Table-1 Spaceflight Sequence Chart

| S.No | Spacecraft | ORBITERS | | | | | | | | | | | | | | | | LANDERS AND ROVERS | | | | | |
|---|---|---|---|---|---|---|---|---|---|---|---|---|---|---|---|---|---|---|---|---|---|---|---|
| | | LNH | SB | SS | LEOI | VS | CRC-1 | OMB | TMTI | CRC-2 | TCM | MA | CRC-3 | CSS | OIB | MOI | HBE | HSJ | PDD | RPD | AD | CRC-4 | LDNG | CRC-5 |
| 1. | 1M No.1 | ● | ⊗ | | | | | | | | | | | | | | | | | | | | | |
| 2. | 1M No.2 | ● | ⊗ | | | | | | | | | | | | | | | | | | | | | |
| 3. | 2MV-4 No.1 | ⊗ | | | | | | | | | | | | | | | | | | | | | | |
| 4. | 2MV-4 No.2 | ● | ● | ● | ● | ● | ● | ⊗ | | | | | | | | | | | | | | | | |
| 5. | 2MV-3 No.1 | ● | ● | ● | ● | ● | ⊗ | | | | | | | | | | | | | | | | | |
| 6. | Mariner 3 | ● | ● | ● | ● | ⊗ | | | | | | | | | | | | | | | | | | |
| 7. | Zond 2 | ● | ● | ● | ● | ● | ● | ● | ● | ⊗ | | | | | | | | | | | | | | |
| 8. | 2M No.521 | ● | ⊗ | ⊗ | | | | | | | | | | | | | | | | | | | | |
| 9. | 2M No.522 | ● | ⊗ | | | | | | | | | | | | | | | | | | | | | |
| 10. | Mariner 8 | ⊗ | | | | | | | | | | | | | | | | | | | | | | |
| 11. | Kosmos 419 | ● | ● | ● | ● | ⊗ | | | | | | | | | | | | | | | | | | |
| 12. | Mars 2 lander | ● | ● | ● | ● | ● | ● | ● | ● | ● | ● | ● | ● | ● | ● | ● | ● | ⊗ | | | | | | |
| 13. | Prop-M | ● | ● | ● | ● | ● | ● | ● | ● | ● | ● | ● | ● | ● | ● | ● | ● | ⊗ | | | | | | |
| 14. | Mars 3 lander | ● | ● | ● | ● | ● | ● | ● | ● | ● | ● | ● | ● | ● | ● | ● | ● | ● | ● | ● | ● | ● | ● | ⊗ |
| 15. | Prop-M | ● | ● | ● | ● | ● | ● | ● | ● | ● | ● | ● | ● | ● | ● | ● | ● | ● | ● | ● | ● | ● | ● | ⊗ |
| 16. | Mars 4 orbiter | ● | ● | ● | ● | ● | ● | ● | ● | ● | ⊗ | | | | | | | | | | | | | |
| 17. | Mars 6 lander | ● | ● | ● | ● | ● | ● | ● | ● | ● | ● | ● | ● | ● | ● | ● | ● | ● | ● | ● | ● | ⊗ | | |
| 18. | Mars 7 lander | ● | ● | ● | ● | ● | ● | ● | ● | ● | ● | ● | ● | ● | ⊗ | | | | | | | | | |
| 19. | Phobos 1 | ● | ● | ● | ● | ● | ● | ● | ● | ⊗ | | | | | | | | | | | | | | |
| 20. | Phobos 2 | ● | ● | ● | ● | ● | ● | ● | ● | ● | ● | ● | ⊗ | | | | | | | | | | | |
| 21. | Mars Observer | ● | ● | ● | ● | ● | ● | ● | ● | ● | ● | ● | ⊗ | | | | | | | | | | | |
| 22. | Mars 96 | ● | ● | ● | ⊗ | | | | | | | | | | | | | | | | | | | |
| 23. | Mars 96 | ● | ● | ● | ⊗ | | | | | | | | | | | | | | | | | | | |
| 24. | Mars 96 | ● | ● | ● | ⊗ | | | | | | | | | | | | | | | | | | | |
| 25. | Nozomi | ● | ● | ● | ● | ● | ● | ● | ● | ● | ⊗ | | | | | | | | | | | | | |
| 26. | MCO | ● | ● | ● | ● | ● | ● | ● | ● | ● | ● | ● | ● | ● | ● | ⊗ | | | | | | | | |
| 27. | MPL | ● | ● | ● | ● | ● | ● | ● | ● | ● | ● | ● | ● | ● | ● | ● | ● | ● | ● | ⊗ | ⊗ | ⊗ | | |
| 28. | Deep Space 2 | ● | ● | ● | ● | ● | ● | ● | ● | ● | ● | ● | ● | ● | ● | ● | ● | ● | ● | ⊗ | ⊗ | ⊗ | | |
| 29. | Beagle 2 lander | ● | ● | ● | ● | ● | ● | ● | ● | ● | ● | ● | ● | ● | ● | ● | ● | ● | ● | ● | ● | ● | ● | ⊗ |
| 30. | Fobos-Grunt | ● | ● | ● | ● | ● | ● | ⊗ | | | | | | | | | | | | | | | | |
| 31. | Fobos-Grunt | ● | ● | ● | ● | ● | ● | ⊗ | | | | | | | | | | | | | | | | |
| 32. | Yinghuo-1 | ● | ● | ● | ● | ● | ● | ⊗ | | | | | | | | | | | | | | | | |
| 33. | Schiaparelli | ● | ● | ● | ● | ● | ● | ● | ● | ● | ● | ● | ● | ● | ● | ● | ● | ● | ● | ● | ● | ⊗ | | |

● Achieved ⊗ Failed

## Table 2. Comparative Summary of unsuccessful Mars Missions

| S.No | Spacecraft | Type | Launch | Launcher | Launcher Type | Country | Issue | Outcomes |
|---|---|---|---|---|---|---|---|---|
| 1. | 1M No.1 | Flyby | 10 Oct 1960 | Molniya | 8K78/L1-4 | USSR | Launch failure | Disintegrated in LEO |
| 2. | 1M No.2 | Flyby | 14 Oct 1960 | Molniya | 8K78/L1-5 | USSR | Failed LEO | Never achieved LEO |
| 3. | 2MV-4 No.1 | Flyby | 24 Oct 1962 | Molniya | 8K78/T-103-15 | USSR | Rocket exploded | Spacecraft destroyed |
| 4. | Mars 1 | Flyby | 01 Nov 1962 | Molniya | 8K78/T103-16 | USSR | Orientation Failure | Lost communication before flyby |
| 5. | 2MV-3 No.1 | Lander | 04 Nov 1962 | Molniya | 8K78/T103-17 | USSR | Failed Ignition | Disintegrated in LEO |
| 6. | Mariner 3 | Flyby | 05 Nov 1964 | Atlas | LV-3 Agena-D | USA | Stage Separation | Lost communication |
| 7. | Zond 2 | Flyby | 30 Nov 1964 | Molniya | 8K78 | USSR | Delayed Solar Panel Deployment | Lost communication |
| 8. | 2M No.521 | Orbiter | 27 Mar 1969 | Proton | K/D UR-500 | USSR | Failed Ignition | Booster exploded destroying the craft |
| 9. | 2M No.522 | Orbiter | 2 Apr 1969 | Proton | K/D UR-500 | USSR | Booster fire accident | Rocket nosedived into the ground |
| 10. | Mariner 8 | Orbiter | 09 May 1971 | Atlas | SLV-3C Centaur D | USA | Failure of Transistor and Guidance Control System | Launch failure |
| 11. | Kosmos 419 | Orbiter | 10 May 1971 | Proton | K/D UR-500 | USSR | Improper stage ignition | Re-entered atmosphere and decayed |
| 12. | Mars 2 | Lander | 19 May 1971 | Proton | K/D UR-500 | USSR | Lander malfunctioned | Crashed on the Martian surface |
| 13. | Prop-M | Rover | 19 May 1971 | Proton | K/D UR-500 | USSR | - | Lost with Mars 2 lander |
| 14. | Mars 3 | Lander | 28 May 1971 | Proton | K/D UR-500 | USSR | Lander's communication system impaired on Mars | Lost communication from ground |
| 15. | Prop-M | Rover | 28 May 171 | Proton | K/D UR-500 | USSR | - | Never deployed from Mars 3 lander |
| 16. | Mars 4 | Orbiter | 21 Jul 1973 | Proton | K/D UR-500 | USSR | Onboard computer failure due to defective transistor | Failed to perform Orbital Insertion |
| 17. | Mars 6 | Lander | 05 Aug 1973 | Proton | K/D UR-500 | USSR | Lander's radio communication system impaired | Lost contact due to Martian rough terrain |
| 18. | Mars 7 | Lander | 09 Aug 1973 | Proton | K/D UR-500 | USSR | Onboard computer failure / failed retrorocket ignition | Failed to enter Martian atmosphere |
| 19. | Phobos 1 | Orbiter | 07 Jul 1988 | Proton | K/D UR-500 | USSR | Wrong Programme commanded | Lost communication from ground |
| 20. | Phobos 2 | Orbiter | 12 Jul 1988 | Proton | K/D UR-500 | USSR | Failure of Orientation System | Lost radio communication |
| 21. | Mars Observer | Orbiter | 25 Sep 1992 | Titan | Titan-III | USA | Spacecraft malfunctioned / short circuit | Lost communication |
| 22. | Mars 96 | Orbiter | 16 Nov 1996 | Proton | K/D-2 UR-500 | USSR | Failed Stage Ignition | Decayed and crashed on earth |
| 23. | Mars 96 | Lander | 16 Nov 1996 | Proton | K/D-2 UR-500 | USSR | Failed Stage Ignition | Decayed and crashed on earth |
| 24. | Mars 96 | Penetrator | 16 Nov 1996 | Proton | K/D-2 UR-500 | USSR | Failed Stage Ignition | Decayed and crashed on earth |
| 25. | Nozomi | Orbiter | 03 Jul 1998 | M/V | MV | JAPAN | Electrical and Communication system impaired | Solar radiation impaired the craft and lost |
| 26. | MCO | Orbiter | 11 Dec 1998 | Delta | II 7425 | USA | Unit conversion software issue (IMU) | Destroyed in upper Mars atmosphere |
| 27. | MPL | Lander | 03 Jan 1999 | Delta | II 7425 | USA | Flight software / Premature engine shutdown | Lost signal and landing is uncertain |
| 28. | Deep Space 2 | Penetrator | 03 Jan 1999 | Delta | II 7425 | USA | - | No signal was received |
| 29. | Beagle 2 | Lander | 02 Jun 2003 | Soyuz | FG/Fregat | USA | Technical and programmatic issue | Lost communication from ground |
| 30. | Fobos-Grunt | Orbiter | 08 Nov 2011 | Zenit | 2M-2FG | USSR | Onboard Computer malfunction and failed ignition | Destroyed during re-entry |
| 31. | Fobos-Grunt | Lander | 08 Nov 2011 | Zenit | 2M-2FG | USSR | Onboard Computer malfunction and failed ignition | Destroyed during re-entry |
| 32. | Yinghuo-1 | Orbiter | 08 Nov 2011 | Zenit | 2M-2FG | CHINA | - | Lost with Fobos-Grunt mission |
| 33. | Schiaparelli | Lander | 14 Mar 2016 | Proton | M/Briz-M | EUROPE | Failure of Inertial Measurement Unit | Crashed on the planetary surface |

Shown in (Fig-4), country (Fig-2), and launch vehicles (Fig-3). And from Fig-2, the duration bars of probes with least are found to be lost with launch vehicles (i.e. few seconds into the flight after launch), whereas Nozomi has the most reliable duration before mission loss. Moreover, from Fig-3, we found Russia had the most probe loss than the United States, Europe, Japan, and China. Similarly, the orbiters encounter more damage than any other type of probes shown in Fig-5. Finally, the launcher Molniya and Proton were accountable for the loss of an excessive number of probes. Furthermore, we have prepared a spaceflight sequence chart representing the mission sequence from launch to the Mars landing in table-1 (Graphical view in Fig. 6) and the overall report is comprehensively cataloged in table-2.

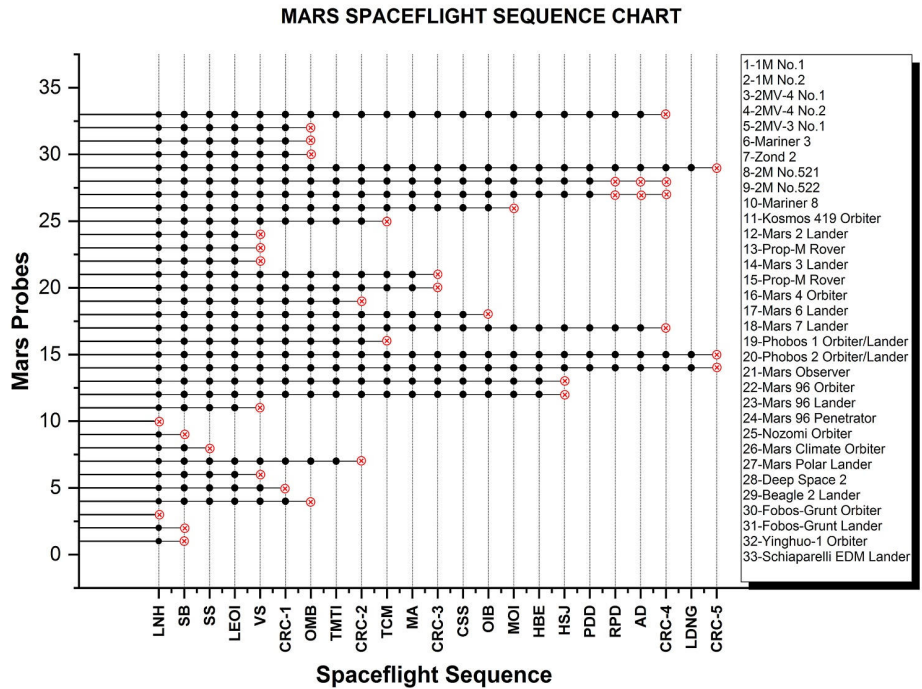

**Fig. 6  Mars Spaceflight Sequence Chart (Graphical View) [78]**

## IX.Conclusion

Our summarized report represents failure records of the entire spacecraft targeted towards Mars. Most of the preceding spacecrafts lost due to obstacles encountered during launch vehicle performance and booster stage ignition that have been enhanced nowadays. But modern spacecrafts attempted after Mars 96 encountered technical issues that have to be taken into consideration for prospects. Detailed analysis of Mars probe failures and possible recommendations were clearly explained in [81]. Insights to future prosperous Mars missions, several investigation articles have been thoroughly analyzed and the root causes for all the unsuccessful Mars crafts have been precisely summarized. Additionally, major issues and their consequences have also been comprehensively tabularized and the mission target achieved during spaceflight in their transit sequence from Earth to Mars has been discussed in detail. Moreover, the proportions of issues encountered by spacecrafts and overall mission duration of spacecrafts have been graphically shown along with a comparative graph depicting the number of failed spacecrafts by country, launch vehicle, and their spacecraft types. This report study was conducted from the perspective of an evident understanding of overall Mars mission probes. Our study is novel and no report was compiled revealing the root causes behind the loss of Mars probes.

## Acknowledgments

The main author Malaya Kumar Biswal would like to extend his sincere thanks and gratitude to his research guide Dr. A. Ramesh Naidu (author of this paper) for his hearted and continued support since the second year of my UG research career.

The authors would like to extend theirs sincere thanks to the general manager Mr. Raajesh Ghoyal, Pulkit Metals Private Ltd, Puducherry for financial assistance for the conference participation.

## Dedication

The main author Malaya Kumar Biswal would like to dedicate this work to his beloved mother late. Mrs. Malathi Biswal, family and friends for their motivational speech and support throughout his life.